\documentclass[10pt,conference]{IEEEtran} 
\pdfoutput=1
\usepackage[english]{babel}
\usepackage{mathptm}
\usepackage{times}
\usepackage{pstricks}
\usepackage{latexsym}
\usepackage[commentsnumbered]{algorithm2e}
\usepackage{subfigure}
\usepackage{graphicx}
\usepackage{framed}
\usepackage[cmex10]{amsmath}
\usepackage{caption}
\usepackage{etoolbox}
\makeatletter
\patchcmd{\maketitle}{\@copyrightspace}{}{}{}
\makeatother

\begin{document}
%
\title{Exploring Task Mappings on Heterogeneous MPSoCs using a Bias-Elitist Genetic Algorithm}

\author{
%
%
\IEEEauthorblockN{Wei Quan$^{\dagger,\ddagger}$}
\IEEEauthorblockA{\\
$^{\dagger}$Informatics Institute\\
University of Amsterdam\\
The Netherlands\\
\{w.quan,a.d.pimentel\}@uva.nl}
\and
\IEEEauthorblockN{Andy D. Pimentel$^{\dagger}$}
\IEEEauthorblockA{\\
$^{\ddagger}$School of Computer Science\\
National University of Defense Technology\\
Hunan, China\\
quanwei02@gmail.com}
}

\maketitle

\begin{abstract}
Exploration of task mappings plays a crucial role in achieving high performance in heterogeneous multi-processor system-on-chip (MPSoC) platforms. The problem of optimally mapping a set of tasks onto a set of given heterogeneous processors for maximal throughput has been known, in general, to be NP-complete. The problem is further exacerbated when multiple applications (i.e., bigger task sets) and the communication between tasks are also considered. Previous research has shown that Genetic Algorithms (GA) typically are a good choice to solve this problem when the solution space is relatively small. However, when the size of the problem space increases, classic genetic algorithms still suffer from the problem of long evolution times. To address this problem, this paper proposes a novel bias-elitist genetic algorithm that is guided by domain-specific heuristics to speed up the evolution process. Experimental results reveal that our proposed algorithm is able to handle large scale task mapping problems and produces high-quality mapping solutions in only a short time period.

\end{abstract}


%

\section{Introduction}
Heterogeneous MPSoC platforms have in recent years received much attention due to their capability of providing good performance and energy consumption trade-offs~\cite{KumarISCA04}. For heterogeneous MPSoC systems, the task mapping problem -- consisting of assigning a set of application tasks to processors and binding communications between tasks to communication channels or memories in the system --  plays a crucial role in achieving high performance. Many heuristic algorithms exist for this task mapping problem, which
can roughly be divided into two categories: the ones that assign one task at a time like Minimum Execution Time (MET) or Minimum Completion Time (MCT)~\cite{Braun01} and the algorithms that map all the tasks at once like Simulated Annealing~\cite{OrsilaThesis} or Genetic Algorithms~\cite{Alexandrescu11}. Comparing these two classes of algorithms, the former category of algorithms usually has lower algorithmic complexity, which means a shorter computing time, but they also produce poorer results.
For the second category of task mapping algorithms, several investigations~\cite{Braun01,Choi12,Page10,Erbas06} have shown that Genetic Algorithms (GA) can consistently generate efficient mapping solutions, also in comparison to alternative heuristic search methods like Simulated Annealing (SA), in a relatively short time period. However, for large problem sizes (i.e., search spaces),  GAs will typically suffer from large computational costs as a significant number of solution evaluations are needed to find good solutions~\cite{Singh13}. Therefore, it is essential to develop effective pruning techniques that can optimize the search process, allowing the design space exploration (DSE) algorithms to explore larger design spaces.

\begin{figure}[t!]
\centering
\subfigure[Makespan versus real performance of the mappings in the mapping space of MJPEG]{
            \label{Fig mjpeg_mp}
            \includegraphics[width=2.6in]{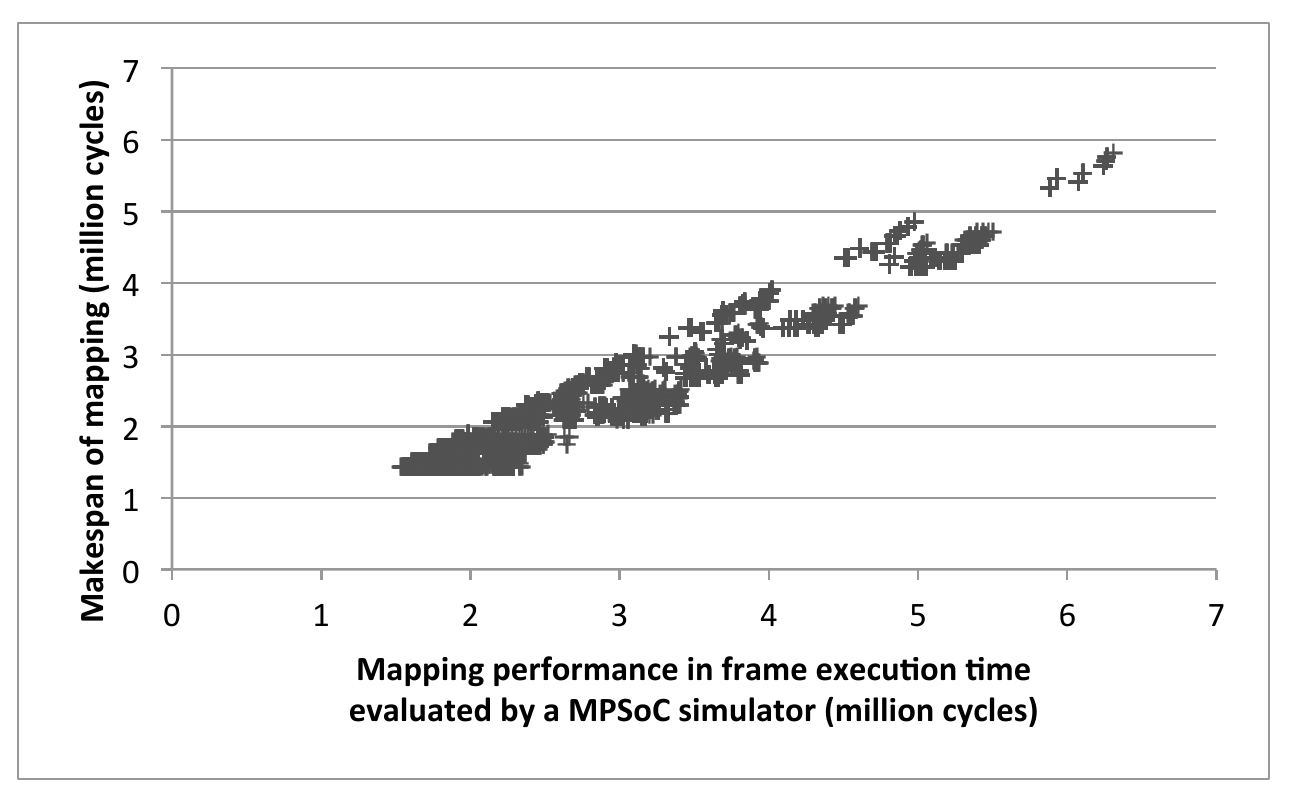}
        }
\subfigure[Processor workload difference versus real performance of mappings with a small makespan value (y-value under 1.5 in the graph of (a)) of MJPEG]{
            \label{Fig mjpeg_wp}
            \includegraphics[width=2.6in]{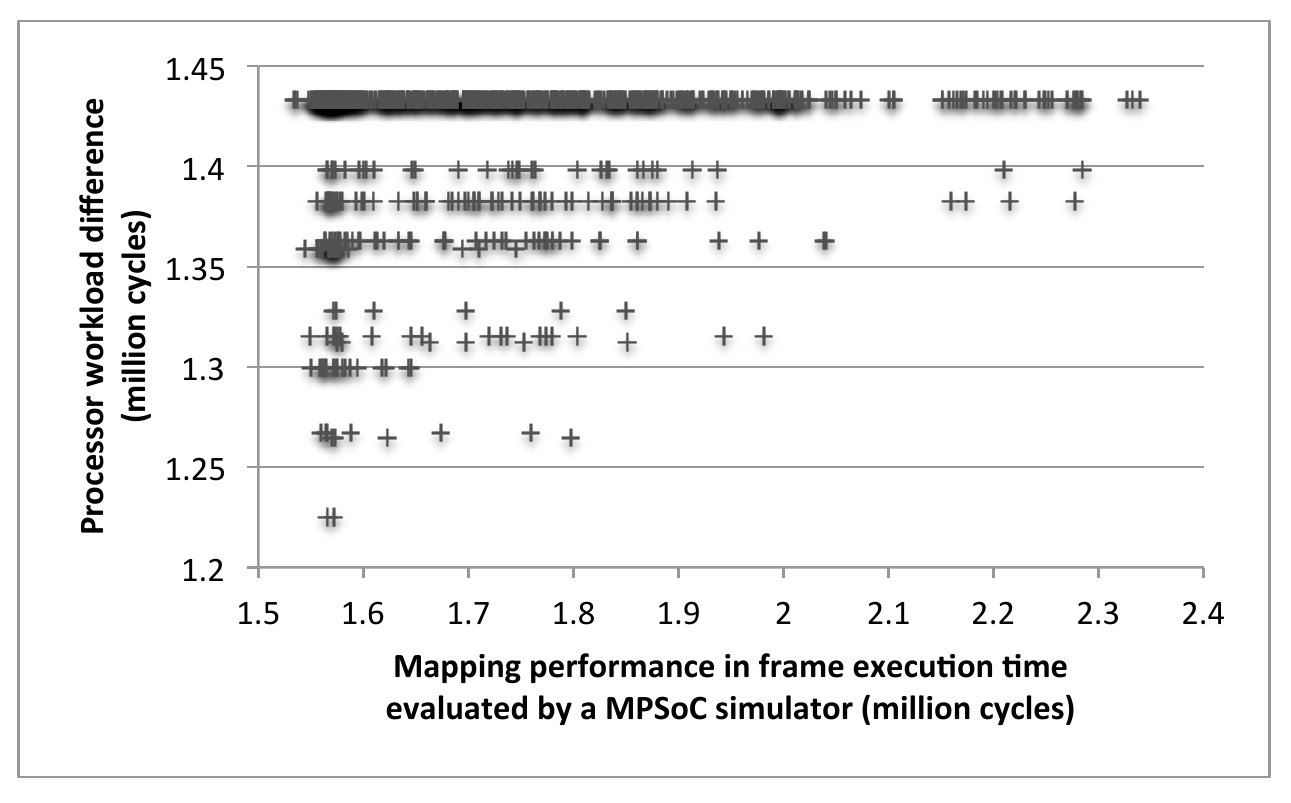}
        }
\caption{Relating real mapping performance to heuristic performance metrics for MJPEG.}
\label{mjpeg}
\vskip -.4cm
\end{figure}

This paper introduces a new GA-based mapping DSE algorithm that allows for effectively pruning the search space in order to reduce the search time. To this end, the algorithm aims at optimizing the genetic operators in the GA that take care of deriving new individuals -- representing design points -- from the old individuals during search iterations.  If the operators can be optimized such that they only generate a small set of chromosomes that has a high probability of containing the optimal or near optimal solutions, then the search time for a good result can be greatly reduced. In this paper, we hypothesize  that such an optimization of the genetic operators is possible through the exploitation of domain knowledge as captured by means of heuristics. To motivate this, please consider the following experiment in which we have exhaustively explored the mapping space of a Motion-JPEG decoder application for a 5-processor heterogeneous MPSoC. Figure~\ref{mjpeg} shows the relationship between the mapping performance as evaluated by a system-level MPSoC simulator and two (analytical) performance heuristics of the same mapping solution, namely the makespan of the mapping and the processor workload imbalance. Although Figure~\ref{Fig mjpeg_mp} indicates that the makespan heuristic cannot predict the mapping performance with high accuracy (i.e., Figure~\ref{Fig mjpeg_mp} does not show a narrow linear line), it clearly shows a linear relationship, and thus a correlation, between mapping performance and makespan. 

Looking more deeply into the mappings with a small makespan, we can see from Figure~\ref{Fig mjpeg_wp} that the mappings with a smaller processor workload imbalance have a higher probability to be a good mapping solution. This means that good results for our mapping problem typically have some common properties such as a small makespan and a workload that is well balanced over processors.
Based on this observation, we propose a novel {\em bias-elitist genetic algorithm} in which the genetic operators have been optimized using application domain knowledge as captured by means of heuristics. 
We will show that our algorithm is able to find high-quality mapping solutions for applications that contain a large number of tasks, and it will do so in much shorter time frames as compared to a range of other well-known algorithms.

The remainder of this paper is organized as follows. Section 2 gives some prerequisites for this paper. Section 3 provides a detailed description of our bias-elitist genetic algorithm. Section 4 introduces the experimental environment and presents the results of our experiments. Section 5 discusses related work, after which Section 6 concludes the paper.

\section{Prerequisites}
\subsection{Target Applications} \label{app_model}
In this paper, we target the multimedia application domain. For this reason, we use the Kahn Process Network (KPN) model of computation~\cite{Kah74} to specify application behaviour since this model of computation fits well to the streaming behaviour of multimedia applications. Consequently, an application can be described as a network of concurrent processes that are interconnected via FIFO channels.  This means that an application can be represented as a directed graph $KPN=(P,F)$ where $P$ is the set of processes (tasks) $p_{i}$ in the application and $f_{ij}\in F$ represents the FIFO channel between two processes $p_{i}$ and $p_{j}$. 
Our goal is to quickly find a task mapping (including the mapping of processes and FIFO communication channels) for the application(s) on the target architecture that maximizes throughput. Here, we consider target applications, or a set of multiple applications that needs to be simultaneously mapped, that contain, in total, a large number of tasks. 

\subsection{Target Architectures} \label{arch_model}
In this work, the target architectures are heterogeneous MPSoC systems in which each processor may have different computational characteristics, making the task mapping problem more complex. 
The architecture can be modeled as a graph $MPSoC=(PE,C)$, where $PE$ is the set of processing elements used in the architecture and $C$ is a multiset of pairs $c_{ij}=(pe_{i},pe_{j}) \in PE \times PE$ representing a communication channel (like Bus, NOC, etc.) between processors $pe_{i}$ and $pe_{j}$. Combining the definition of application and architecture models, the computation cost of task (process) $p_{i}$ on processing element $pe_{j}$ is expressed as $T_{i}^j$ and the communication cost between tasks $p_{i}$ and $p_{j}$ on channel $c_{xy}$ is $C_{ij}^{c_{xy}}$. Here, $c_{xy}$ represents the communication channel between processor $pe_{x}$ and $pe_{y}$ where tasks $p_{i}$ and $p_{j}$ are mapped onto respectively. In this paper, we restrict ourselves to MPSoC systems with shared memory. This implies that a FIFO channel can either be mapped onto the internal local memory of a processor (when two communicating processes are mapped onto the same processor) or onto shared memory. In other words, if $pe_{x}$ equals to $pe_{y}$, the communication channel $c_{xy}$ is the processor's local memory, otherwise it is the shared memory. 

\subsection{Simulation Framework}
\begin{figure}[!t]
\centering
\includegraphics[width=2.6in]{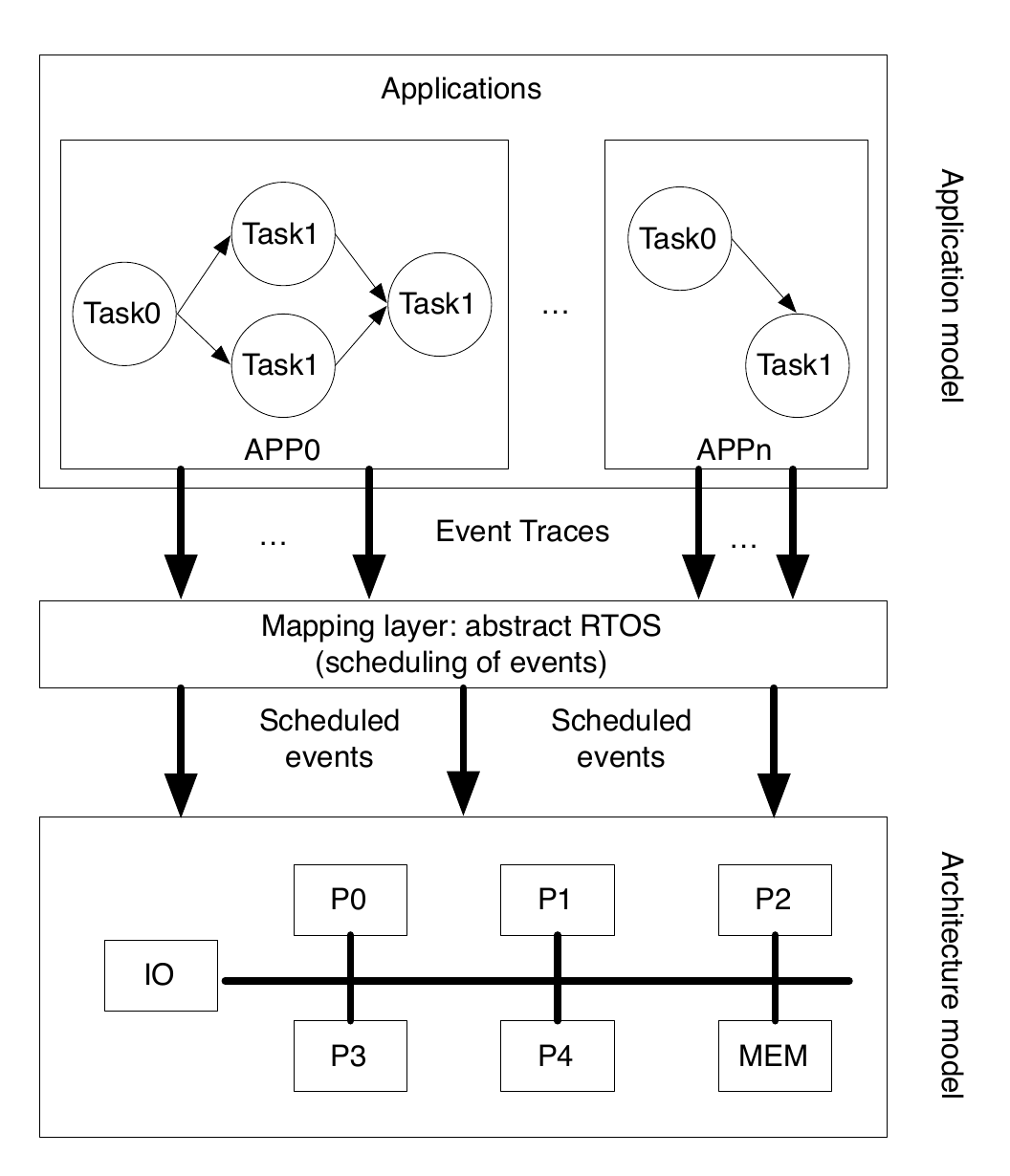}
\caption{Sesame Framework.}
\label{figure2}
\vskip -.4cm
\end{figure}

In our work, we deploy the open-source Sesame system-level MPSoC simulator \cite{PimentelEP06} to evaluate the fitness of mappings. 
The Sesame modeling and simulation environment, which is illustrated in Figure~\ref{figure2}, facilitates efficient performance analysis of embedded (media) systems architectures. It recognizes separate application and architecture models, where an application model describes the functional behavior of an application and the architecture model defines architecture resources and captures their performance constraints. After explicitly mapping an application model onto an architecture model, they are co-simulated via trace-driven simulation. This allows for evaluation of the system performance of a particular application, underlying architecture, and mapping.

Although a Sesame-based simulation of each individual mapping to evaluate its fitness only takes a few seconds, the total evaluation time for solving large task mapping problems may grow to an unacceptable level. This underlines the need of reducing the evolution time of the genetic algorithm that searches the mapping space.

\section{Bias-elitist Genetic Algorithm}
Our bias-elitist GA combines a form of elitism as found in classic elitist GAs with the concept of a domain knowledge guided GA such as from \cite{Alexandrescu11}. It tries to find a task mapping for the target application(s) on a heterogeneous MPSoC system with the objective of maximizing the throughput. The details of our domain knowledge guided GA will be explained in the following subsections.

\subsection{Encoding and initial population}
Each mapping solution is encoded as a string of integers. The tasks of the target application(s) are arranged in the chromosome according to the topological order in the application KPN. Each gene in the chromosome represents a unique identifier of the processors in the MPSoC system (i.e., denoting the processor the task is mapped on). As our target architecture is a shared-memory heterogeneous MPSoC system, communication is performed either via the internal local memory on each processor or the shared memory. 
Consequently, we do not need to explicitly encode the mapping of the FIFO channels in a  chromosome which greatly simplifies the mapping problem but without loss of generality serves our purpose of showing the effectivity of our GA. However, we do want to stress that our GA can easily be extended to include explicit channel mappings, such as e.g. in \cite{Erbas06}.

In our GA, the chromosomes in the initial population are randomly generated. Moreover, we limit the size of the initial population as well as the size of the set of generated individuals during each evolution generation of the GA to reduce the simulation time. 

\subsection{Fitness Function}
The fitness function is defined for measuring the quality of solutions. 
In our task mapping problem, we not only need to optimize the makespan of application tasks like in a general task mapping problem (mapping independent tasks) but also the communication between tasks. Here, the resource contention and task communication should be carefully considered in the exploration. As analytical fitness evaluation approaches typically are not capable of accurately capturing such aspects, we deploy the open-source Sesame system-level MPSoC simulator \cite{PimentelEP06} to accurately evaluate the fitness of each chromosome, i.e., mapping, in the population.

\subsection{Selection}
During each successive generation of the GA, a proportion of the existing population is selected to breed a new generation. Individual solutions are selected through a fitness-based selection process, where fitter solutions are typically more likely to be selected. 
Our algorithm uses a roulette wheel selection method in which the best chromosomes are more likely to be selected but the poorer chromosomes also have a small chance to be picked. We should note that in this paper we only consider the single-objective optimization problem. For multi-objective optimization problems (e.g., simultaneously optimizing performance and power consumption), the well-known selection approaches from the NSGA-II or SPEA-II GAs would be good options.

To control the population size in each generation, we use a strategy in which the best chromosome from the current population and $n-1$ chromosomes from the newly generated population are selected as the $n$ survived individuals to breed the next new generation. The rationale behind this is that we aim at increasing the diversity of chromosomes in the mapping space that will be searched by keeping as few as possible old individuals in the new population. Therefore, in contrast to a general elitist GA, where the elitists in each generation will survive in the next generation, our GA only preserves the best individual in each generation. It is an extreme instance of an elitist GA. The fact that we refer to our GA as a {\em bias}-elitist GA will be explained in the next section. 

\subsection{Genetic operators}
To generate a new generation from the selected chromosomes, two genetic operators -- crossover and mutation -- are applied. In our algorithm, we have improved the mutation operator so that the algorithm can more quickly find better solutions. For the crossover operator, which produces a new pair of chromosomes from a selected pair of chromosomes, we apply a standard one-point crossover. We have chosen this operator because it is simple and produces similar results compared with other crossover operators.


\begin{algorithm}[t]
\LinesNumbered
\renewcommand{\arraystretch}{1.0}
\caption{Heuristic guided mutation}
\label{algorithm2}

\SetKwInOut{Input}{input}\SetKwInOut{Output}{output}
\begin{scriptsize}
\Input{$C$ (old chromosome)}
\Output{$C^*$ (new chromosome)}
$PU$ = $pusage(C)$\;
$x$ = index of processor with $max(PU)$\;
\For{task $p_i$ mapped onto processor $pe_x$}{
   \For{processor $pe_y$ different than $pe_x$}{
       $C'$ = migrate $p_i$ from processor $pe_x$ to $pe_y$\;
       $PU'$ = $pusage(C')$\;
       \If{$max(PU') <= max(PU)$}{
           $MBF.append(B_i^{xy})$\;
       }
   }
}
\eIf{array $MBF$ is not empty}{
   $p_k, pe_k$ = task and target processor with maximal\
    migration benefit ($max(MBF)$)\;
   $C^*$ = migrate $p_k$ from processor $pe_x$ to $pe_k$\;
   goto step 1, start with the new mapping $C^*$\;
}
{
   \If{no new mapping found in the previous steps}{ 
      \For{processor $pe_y$ different than $pe_x$}{
          $C'$ = switch the tasks mapped onto $pe_x$ and $pe_y$\;
          $PU'$ = $pusage(C')$\;
          \If{$max(PU') <= max(PU)$}{
              $SBF.append(max(PU'))$\;
          }
      }
      \eIf{array $SBF$ is not empty}{
         $pe_k$ = processor with $min(SBF)$\;
         $C^*$ = switch the tasks mapped onto $pe_x$ and $pe_k$\;
      }
      {
         shuffle the order of tasks in chromosome\;
         $C^*$ = generate new mapping using the MCT\
         algorithm based on the shuffled task order\;
       }
    }
}
\textbf{return} $C^*$\
\end{scriptsize}
\end{algorithm}

The mutation operator is an essential part of our GA. It allows the GA to search new areas in the solution space. There are various methods of implementing the mutation operator, and the easiest one is the random mutation where a random task is re-assigned to a random processor. In our algorithm, we deploy a heuristic guided mutation operator that optimizes the mappings using domain knowledge. More specifically, the mutation operator considers the affinity of tasks with respect to processors, the communication cost between tasks, and the differences of processor workloads. The details of our mutation operator are outlined in Algorithm~\ref{algorithm2}. By applying the mutation operation, a new chromosome will be derived through one of the following three approaches: task migration (lines 1-15), processor switching (lines 18-27) or a Minimum Completion Time (MCT) algorithm (lines 29-30). 

At the beginning of the mutation, the task migration method will be used to find a new chromosome based on the input chromosome. In this process, the usage of each processor $U_k$ under a given task mapping is calculated by equation~\ref{equ1} in the function at line 1. Lines 3-11 of Algorithm~\ref{algorithm2} try to find a task (among the tasks mapped onto the most heavily loaded processor) that has a maximal "migration benefit" under the condition of line 7. This task migration benefit, with regard to task $p_i$ migrated from processor $pe_x$ to $pe_y$, is labeled as $B_i^{xy}$. It is calculated by equation~\ref{equ2} where $M_i^x$ and $M_i^y$ represent the cost of $p_i$ on $pe_x$ and $pe_y$ respectively. Here, the cost not only considers the task computation time but also the accumulated communication costs of the task in question. 
\begin{equation}
\label{equ1}
U_{k} = \sum_{p_{i} \mapsto pe_{k}, p_{j} \mapsto pe_{y}}{(T_{i}^k+C_{ij}^{c_{ky}})} \\[0.05mm]
\end{equation}
where $a \mapsto b$ implies that $a$ is mapped onto $b$.

\begin{subequations}\label{equ2}
\begin{equation}
B_i^{xy} = M_i^x - M_i^y
\end{equation}
\begin{equation}
M_i^t = T_{i}^t+\sum_{\substack{f_{ij} \in F, c_{tk} \in C \\ p_i \mapsto pe_t, p_j \mapsto pe_k, f_{ij} \mapsto c_{tk}}}{C_{ij}^{c_{tk}}} 
\end{equation}
\end{subequations}

If a task can be found for migration after the steps in lines 3-11, lines 13-14 in Algorithm~\ref{algorithm2} will generate a new mapping by migrating this task to the corresponding target processor. Subsequently, the above process is repeated -- using the new mapping as input --  until no new mapping can be found anymore.

However, if the above task migration approach cannot find a new chromosome, then the processor switching method will be applied to the input chromosome. As shown in lines 18-27 in Algorithm~\ref{algorithm2}, the new chromosome will be generated by exchanging the tasks mapped onto the heaviest loaded processor with the tasks mapped onto the processor which satisfies the conditions on line 21 and line 26 (the processor that will maximally reduce the value of $max(PU)$ by processor switching).

In the case that no new mapping can be found by using any of the two previous approaches, a heuristic-based random mutation operator will be applied. A totally new chromosome, which means all the genes in the chromosome are different from the ones in the input chromosome, might be generated in this approach. The heuristic used for generating a new chromosome is the Minimum Completion Time (MCT) algorithm. The MCT algorithm assigns each task, in arbitrary order, to the processor with the minimum expected completion time for that task~\cite{Armstrong98}. 
Different task assignment orders will produce different mapping results. Therefore, each time before generating a new chromosome using MCT, the task order in the chromosome is shuffled. Consequently, different well-balanced chromosomes will be added to the new population of our GA. This helps our GA to explore the mapping space with more gene diversity and prevents our GA from getting stuck in a local minimum.

\begin{figure}[!t]
\centering
\includegraphics[width=3.0in]{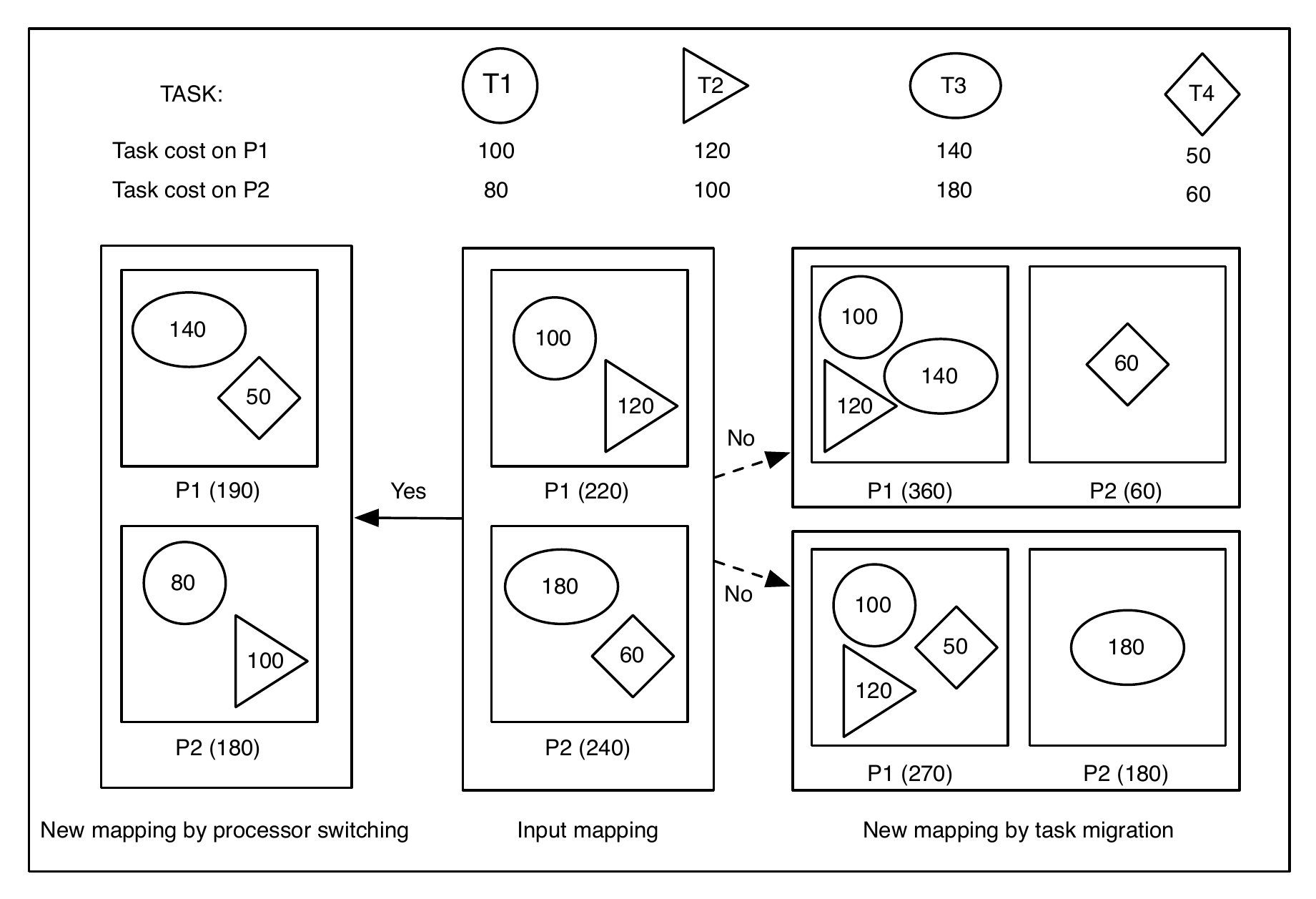}
\caption{A simple example of improving the mapping makespan using processor switching.}
\label{p_switch}
\vskip -.4cm
\end{figure}

A new chromosome generated by our mutation operator has a bias towards design points that are makespan optimized and/or workload balanced. This explains the name {\em bias-elitist GA}. The task migration approach can optimize both the makespan and the processor workload variation. However, the processor switching approach is supplementary to the task migration approach for optimizing the mapping makespan in situations such as illustrated in Figure~\ref{p_switch}. When a chromosome is selected for mutation, Algorithm~\ref{algorithm2} will first try to optimize the mapping makespan and the processor workload balance using the task migration method. However, if this does not succeed (e.g., when the input chromosome already represents a well-balanced system workload), then the processor switching method will be applied to the input chromosome to further improve the mapping makespan. If no improved mapping can be derived from the input chromosome using either method, then the MCT algorithm will be used to generate a new well-balanced mapping.
By applying this domain knowledge guided mutation operator, the search space of our GA will be pruned to only those mappings with a high likelihood of being high-quality mappings. As we will show in the experimental section, this results in a much more efficient and effective search algorithm that allows for producing good mapping solutions in relatively short evolution times.

\subsection{Termination}
With respect to the stopping conditions for our GA, two conditions are used: (1) if the best solution has not changed after a pre-defined number of generations, then our GA will terminate automatically and (2) a maximum number of generations is adopted to guarantee that the evolution process will stop. Our bias-elitist GA aims at reducing the required (maximum) number of iterations as much as possible while still yielding good solutions. The above termination conditions are also applied to the other GAs that are studied in our experiments in the next section.

\section{Experiments}\label{sec:exp}
For our experiments, we have selected a real multi-media application to investigate various aspects of our GA: a MP3 decoder consisting of 27 application processes (tasks).
The target architecture considered in our experiments consists of 5 heterogeneous processors and 1 IO processor (for IO tasks). These processors are connected via a bus to a shared memory. In the MP3 task mapping problem, the total number of possible mapping solutions is $2.98*10^{17}$. Our Bias-Elitist Genetic algorithm (BEG) and several other algorithms will be used to explore this vast solution space to find the (near) optimal mapping with the objective to maximize the throughput. As our BEG algorithm is not limited to only solving the mapping problem for single applications, a multi-application mapping case -- considering a Motion-JPEG encoder and Sobel filter for edge detection in addition to the MP3 decoder -- is studied as well. There, we consider the maximization of system throughput when multiple applications are active simultaneously. The total number of possible mapping solutions in this multi-application mapping problem is $2.91*10^{24}$. 

\begin{table}[!t]
\begin{center}
\centering
\caption{Parameters of genetic algorithms} 
\label{table1}
\begin{scriptsize}
\begin{tabular}{|c|c|c|c|c|}
\hline
\textbf{Parameter} & \multicolumn{2}{c|}{\textbf{Experiment 1}} & \textbf{Experiment 2}  & \textbf{Experiment 3} \\
\cline{2-5}
& EG/GA3SM & BEG & all GAs & all GAs \\
\hline
Initial pop. size & 128 & 8 & 8 & 128 \\
\hline
Generation pop. size & 128 & 8 & 8 & 128 \\
\hline
Crossover prob. & 0.7 & 0.7 & 0.7 & 0.7 \\
\hline
Mutation prob. & 0.8 & 0.8 & 0.8 & 0.8 \\
\hline
Max. \# of generations & 2048 & 128 & 128 & 128 \\
\hline
\end{tabular}
\end{scriptsize}
\end{center}
\vskip -.6cm
\end{table}

\subsection{Single-application Task Mapping}
For the purpose of comparison, three other mapping algorithms are studied as well: a general Elitist Genetic (EG) algorithm~\cite{Dey02}, a Genetic Algorithm with a 3-Step Mutation (GA3SM)~\cite{Alexandrescu11} and Output-Rate Balancing (ORB) \cite{JCII11} which aims at balancing the computation and communication load of each processor. For the genetic algorithms (BEG, EG and GA3SM), the parameters in the experiments of single-application task mapping are listed in Table~\ref{table1}. The parameters of each GA are optimized for each experiment. Notice that the parameter of mutation probability used in our experiments is a chromosome-level concept\footnote{Chromosome-level mutation probability: the likelihood of mutating a particular chromosome.}. It differs from the mutation probability used in typical GAs which is considered at the gene level\footnote{Gene-level mutation probability: the likelihood of mutating each gene (bit) of a chromosome in mutation.} and is usually small ($<0.1$). In our experiments, the gene-level mutation probability only exists in the EG algorithm and its value is 0.05. For the purpose of a fair comparison, the same randomly generated initial population is provided to the EG and BEG algorithms. For the GA3SM algorithm, the initial population is derived by replacing the worst individual in the randomly generated initial population with the result of the Min-Min heuristic. This is according to the original GA3SM algorithm. The results of all experiments have been averaged over 10 execution runs to deal with the stochastic behaviour of the GAs. For all experiments, we have used a PC with a 2.93GHz Intel Core i7 CPU.

\begin{figure}[t!]
\centering
\includegraphics[width=2.6in]{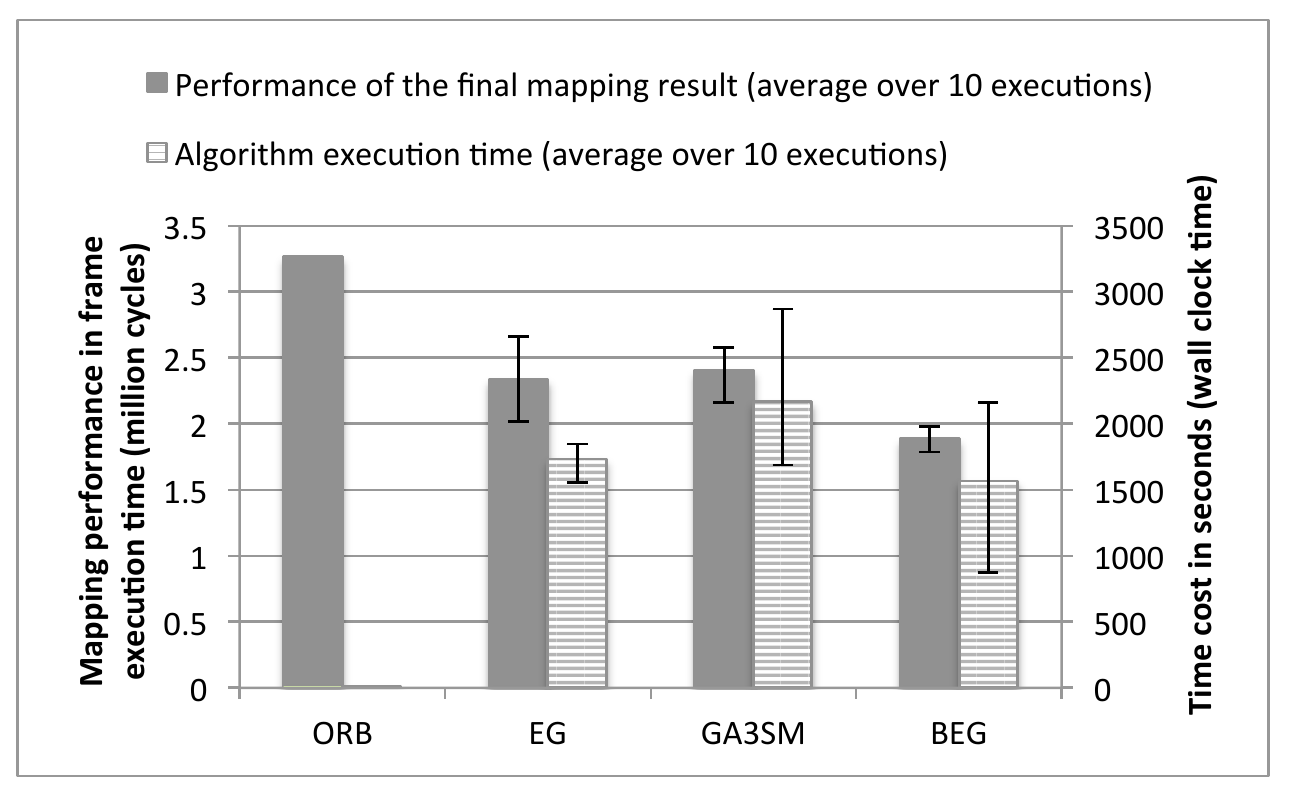}
\caption{The quality of final mapping and search performance for different algorithms.}
\label{figure3}
\vskip -.4cm
\end{figure}

In the first experiment, the original EG and GA3SM algorithms are compared with our BEG algorithm. This means that EG and GA3SM use their own analytical fitness functions to evaluate the fitness value of each chromosome. However, for our BEG algorithm, the fitness value of each chromosome is evaluated using the Sesame simulator. For the BEG algorithm, we have deliberately chosen a small population and generation size to keep the computational costs of the search as low as possible. Figure~\ref{figure3} shows the quality (frame execution time) of the produced mapping solution and the average execution cost for each algorithm (averaged over 10 runs). Here, the ORB algorithm is only executed once as it does not have any stochastic behaviour like the GAs. From this figure, we can clearly see that the EG and GA3SM algorithms cannot generate final mappings as good as the BEG algorithm. Moreover, on average, the EG and GA3SM algorithms also spend more time on finding their final solution.
Apparently, the heuristic ORB method takes the least time to get a final mapping solution which is negligible compared with the execution times of the other algorithms. However, the final mapping derived by ORB is much worse than the ones generated by the GAs.

As the analytic fitness functions used in the original EG and GA3SM algorithms are less accurate than the Sesame-based evaluations in our BEG algorithm, we have also adapted the EG and GA3SM algorithms to use  Sesame to evaluate the fitness of chromosomes.  These Sesame-based EG and GA3SM algorithms are used in the remainder of the experiments.
In the second experiment, we compare our BEG algorithm to the (Sesame-based) EG and GA3SM  algorithms on three aspects: (1) the quality (frame execution time) of the final mapping solution, (2) the algorithm execution time and (3) the convergence behavior of an algorithm. 

\begin{table}[!b]
\vskip -.5cm
\begin{center}
\begin{small}
\centering
\caption{Comparison of final mapping quality in Frame Execution Time (cycles, the smaller the better) and algorithm execution cost (seconds) of GAs with small population size.}  {
\label{table2}
\begin{tabular}{|c|c|c|c|}
\hline
 & EG &  GA3SM & BEG\\\hline
\hline
\textbf{Max. FET}   & 2342502 & 2218522 & 1979684\\
\hline
\textbf{Min. FET} & 2022538 & 1911142 & 1784318\\
\hline
\textbf{Average FET} & 2197809 & 2064753 & 1885810\\\hline\hline
\textbf{Max. cost} & 4897 & 3074 & 2160 \\\hline
\textbf{Min. cost} &  2145 & 1637 & 875 \\\hline
\textbf{Average cost} & 3217& 2245 & 1567 \\\hline
\end{tabular}}
\end{small}
\end{center}
\end{table}



\begin{figure}[!t]
\centering
\includegraphics[width=2.6in]{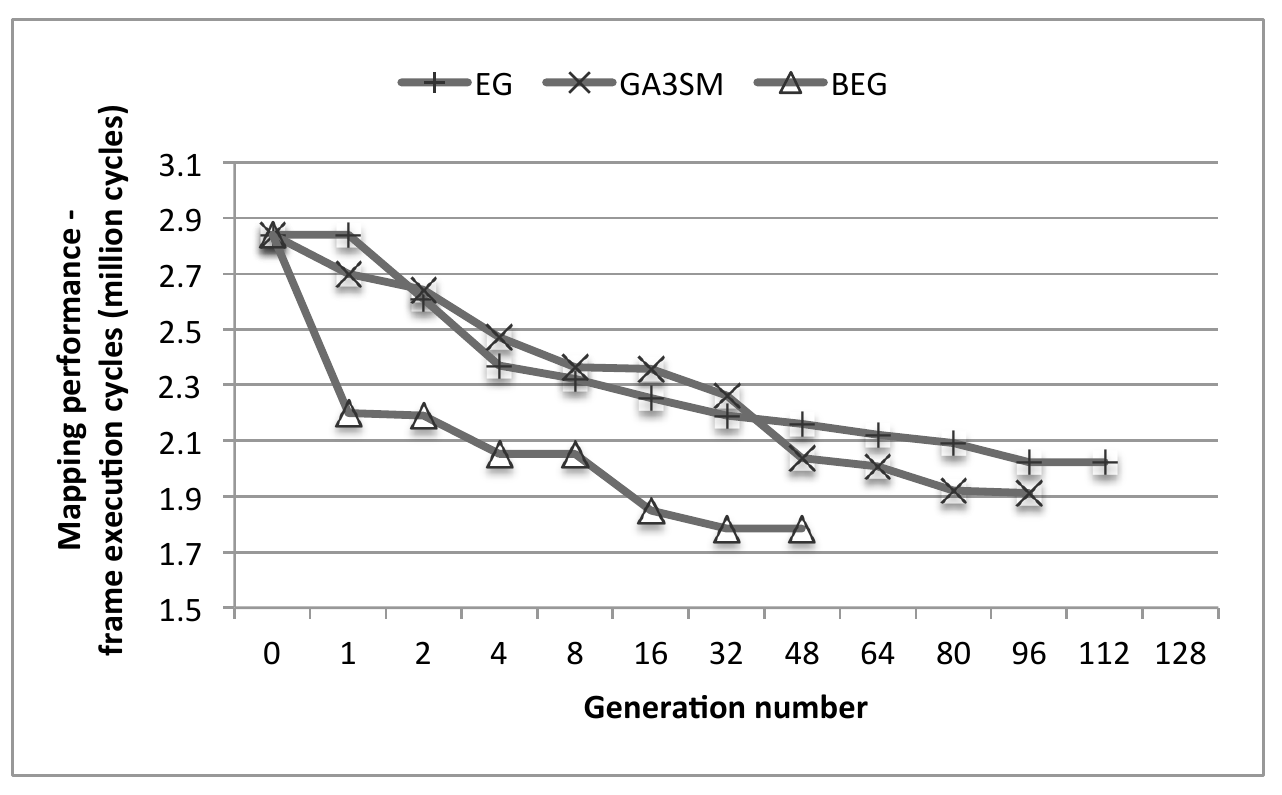}
\caption{The convergence behavior of each GA with a small population size.}
\label{figure4}
\vskip -.4cm
\end{figure}

Table~\ref{table2} shows the quality of the final mappings derived from the different algorithms as well as the algorithm execution cost of the search algorithms. From Table~\ref{table2}, we can see that our BEG algorithm can produce much better solutions than the other GAs. Our BEG algorithm also takes less time to find the final mapping solution as compared to the other two algorithms. The reason for this is that our BEG algorithm can converge much faster than the other two GAs, as shown in Figure~\ref{figure4}. This graph shows the convergence behavior of the execution run for each algorithm that produced the best final solution out of 10 runs.  Considering all 10 runs, our BEG algorithm generated the final mapping solution between the 8th search generation (corresponding to minimal time cost in Table~\ref{table2}) and the 35th generation (corresponding to the maximal time cost in Table~\ref{table2}). For the EG and GA3SM algorithms, however, the final mapping solutions were found between the 9th -- 95th and 12th -- 82th search iterations respectively.


\begin{table}[!b]
\vskip -.4cm
\begin{small}
\begin{center}
\caption{Comparison of final mapping quality in Frame Execution Time (cycles) and algorithm execution cost (seconds) of GAs with large population size.}  {
\label{table4}
\begin{tabular}{|c|c|c|c|}
\hline
 & EG &  GA3SM & BEG\\\hline
\hline
\textbf{Max. FET}   & 2030686 & 1953746 & 1821842\\
\hline
\textbf{Min. FET} & 1862988 & 1768808 & 1762048\\
\hline
\textbf{Average FET} & 1943892 & 1847738 & 1779620\\\hline\hline
\textbf{Max. cost} & 42729 & 55002 & 47824 \\\hline
\textbf{Min. cost} &  21342 & 27814 & 29778 \\\hline
\textbf{Average cost} & 33381& 43274 & 39496 \\\hline
\end{tabular}}
\end{center}
\end{small}
\end{table}



\begin{figure}[t]
\centering
\includegraphics[width=2.6in]{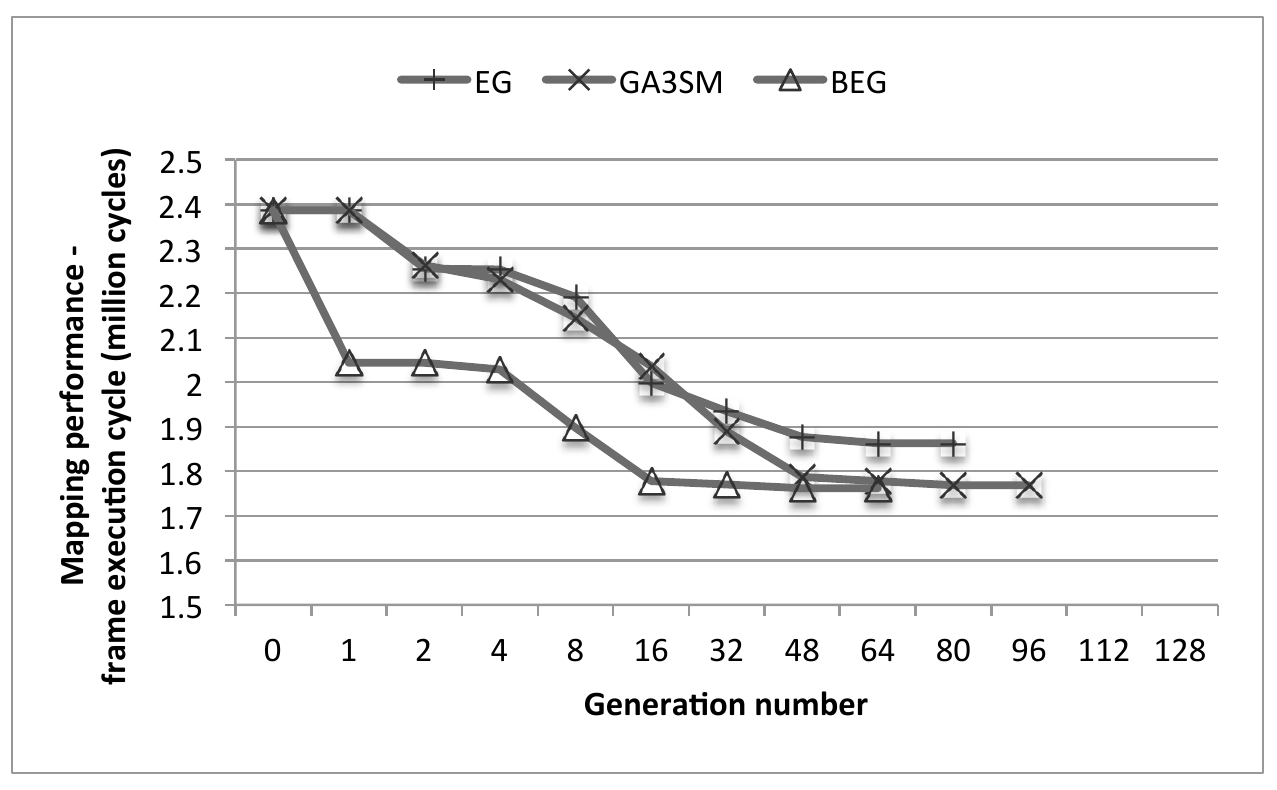}
\caption{The convergence behavior of each GA with a large population size.}
\label{figure5}
\vskip -.4cm
\end{figure}

In the third experiment, we studied the behaviour of each GA using a larger search space by increasing the population size. 
The results are shown in Table~\ref{table4} and Figure~\ref{figure5}.
As shown in Table~\ref{table4}, our BEG algorithm again outperforms the other algorithms with respect to the quality of the final mapping solution. Compared with the second experiment, each algorithm produces better mapping results. More specifically,
the EG, GA3SM and BEG algorithms improve the mapping quality in frame execution time of the best mapping solution (Min. FET in Table~\ref{table2} and Table~\ref{table4}) by 7.9\%, 7.4\% and 1.2\%, respectively. However, these mapping solution improvements come at the expense of a much higher exploration time. The corresponding search times of EG, GA3SM and BEG  increase by 7.1, 15.4 and 45.8 times, respectively. To provide more insight in the large increase of search time for our BEG algorithm,  Figure~\ref{figure5} again shows the convergence behavior. It shows that the BEG algorithm already produces a mapping solution after only 16 generations that is close to the final mapping solution in terms of quality. However, it takes nearly 40 more generations to derive a slightly better final result.
Considering this and the previous experiment together, we can see that our BEG algorithm always yields the best solutions, and on top of this, it can already find a good mapping result in a relatively short time by reducing the population size. The EG and GA3SM algorithms, on the other hand, require algorithm execution times that are about an order of magnitude higher to find similar good mapping results as our algorithm.


\subsection{Multi-application Task Mapping}

In this experiment (the fourth experiment), we investigate our BEG algorithm by solving a multi-application task mapping problem in which the MP3 decoder (27 application tasks), a Motion-JPEG encoder (8 application tasks) and a Sobel filter for edge detection  (6 application tasks) will be mapped onto our previously described target platform. The quality of the final mapping (total execution time) and the algorithm execution cost are compared again for the BEG, EG and GA3SM algorithms. The parameters for each algorithm are the same as in the second experiment from the previous section (see Table~\ref{table1}). The experimental results are shown in Table~\ref{table6}. As can be seen from the results, our BEG algorithm again produces better solutions in a much shorter time frame than the other two GAs.

\begin{table}[!b]
\begin{small}
\begin{center}
\caption{Comparison of final mapping quality in Total Execution Time (cycles) and algorithm execution cost (seconds) of GAs for solving the multi-application mapping problem.}  {
\label{table6}
\begin{tabular}{|c|c|c|c|}
\hline
 & EG &  GA3SM & BEG\\\hline
\hline
\textbf{Max. TET}   & 13200633 & 11857216 & 11288893\\
\hline
\textbf{Min. TET} & 10831581 & 10705250 & 9193485\\
\hline
\textbf{Average TET} & 11804280 & 11209070 & 10120488\\\hline\hline
\textbf{Max. cost} & 7965 & 6174 & 3331 \\\hline
\textbf{Min. cost} &  3844 & 2203 & 1499 \\\hline
\textbf{Average cost} & 6314& 4236 & 2499 \\\hline
\end{tabular}}
\end{center}
\end{small}
\end{table}

\begin{table}[t]
\begin{center}
\centering
\caption{Parameters of genetic algorithms} 
\label{table7}
\begin{scriptsize}
\begin{tabular}{|c|c|c|}
\hline
\textbf{Parameter} & \textbf{Experiment 5} & \textbf{Experiment 6}  \\
\cline{2-3}
&  BEG & BEG  \\
\hline
Initial pop. size & 8 & 8  \\
\hline
Generation pop. size & 8 & 8 \\
\hline
Crossover prob. & 0.7 & 0.7 \\
\hline
Mutation prob. & 0.8 & 0.1--1.0 \\
\hline
Max. \# of generations & 128 & 128 \\
\hline
\end{tabular}
\end{scriptsize}
\end{center}
\end{table}

\begin{figure}[b!]
\centering
\subfigure[Performance of final mappings (10 executions) generated by the BEG algorithm using with different operators]{
            \label{Fig poperator}
            \includegraphics[width=2.6in]{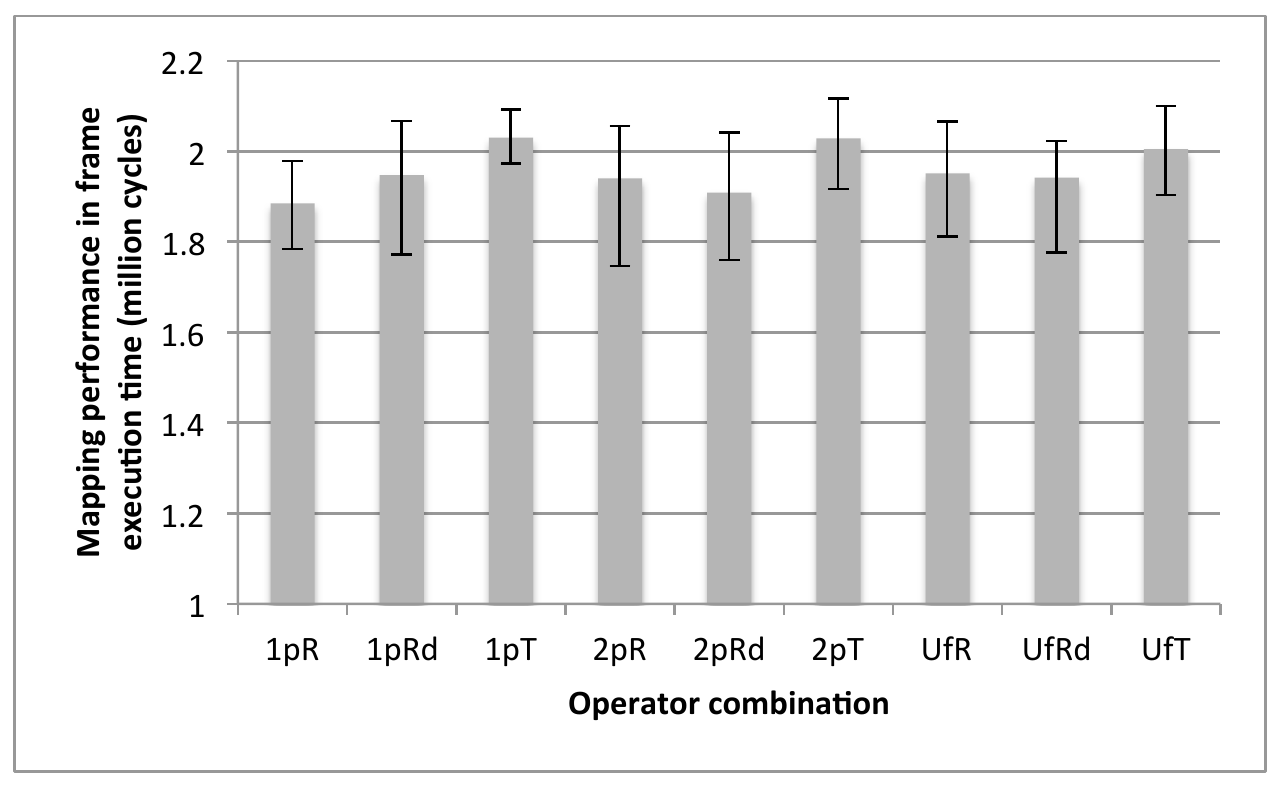}
        }
\subfigure[Algorithm execution time cost (10 executions) of the BEG algorithm with different operators]{
            \label{Fig coperator}
            \includegraphics[width=2.6in]{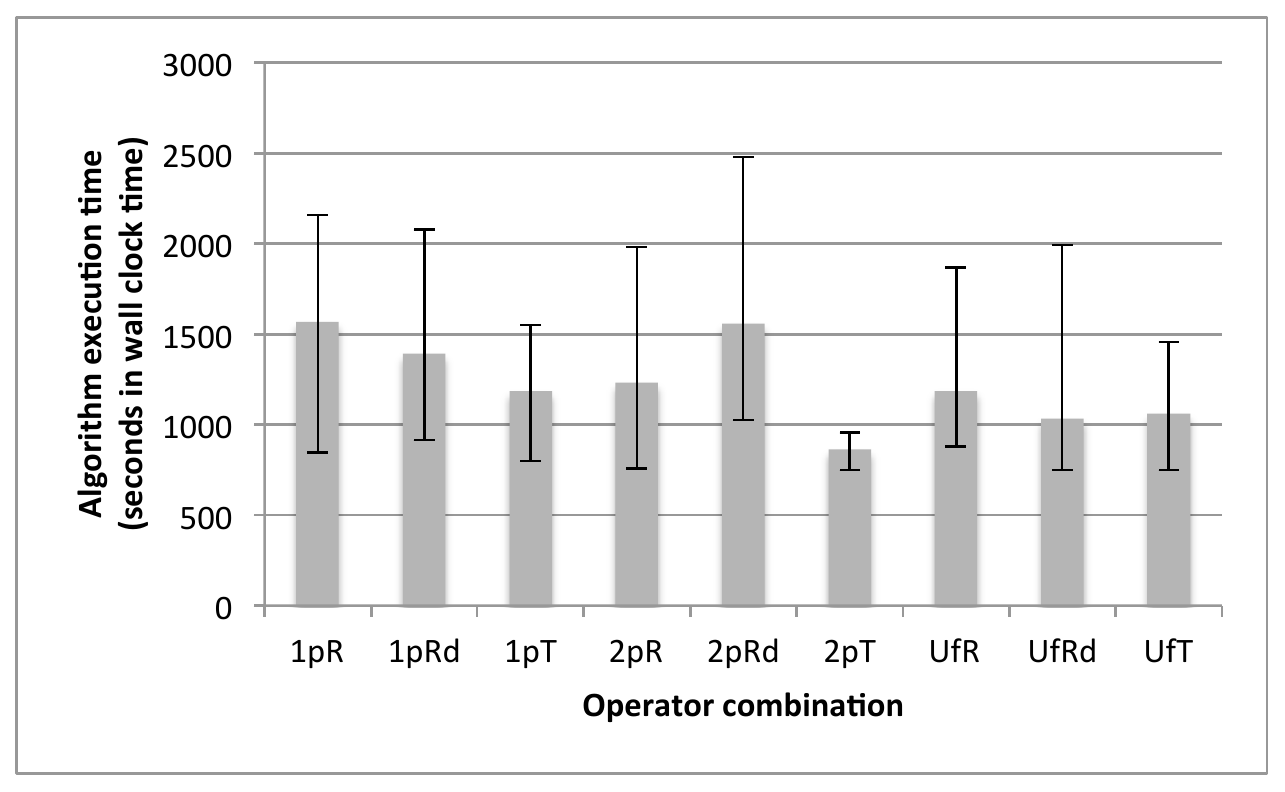}
        }
\caption{Comparison of final mapping performance and algorithm execution time cost of the BEG algorithm with different operators.}
\label{figure6}
\end{figure}

\subsection{Sensitivity to BEG implementation choices}

\begin{figure}[b!]
\centering
\subfigure[Performance of final mappings (10 executions) generated by the BEG algorithm with different mutation probabilities]{
            \label{Fig pmutation}
            \includegraphics[width=2.6in]{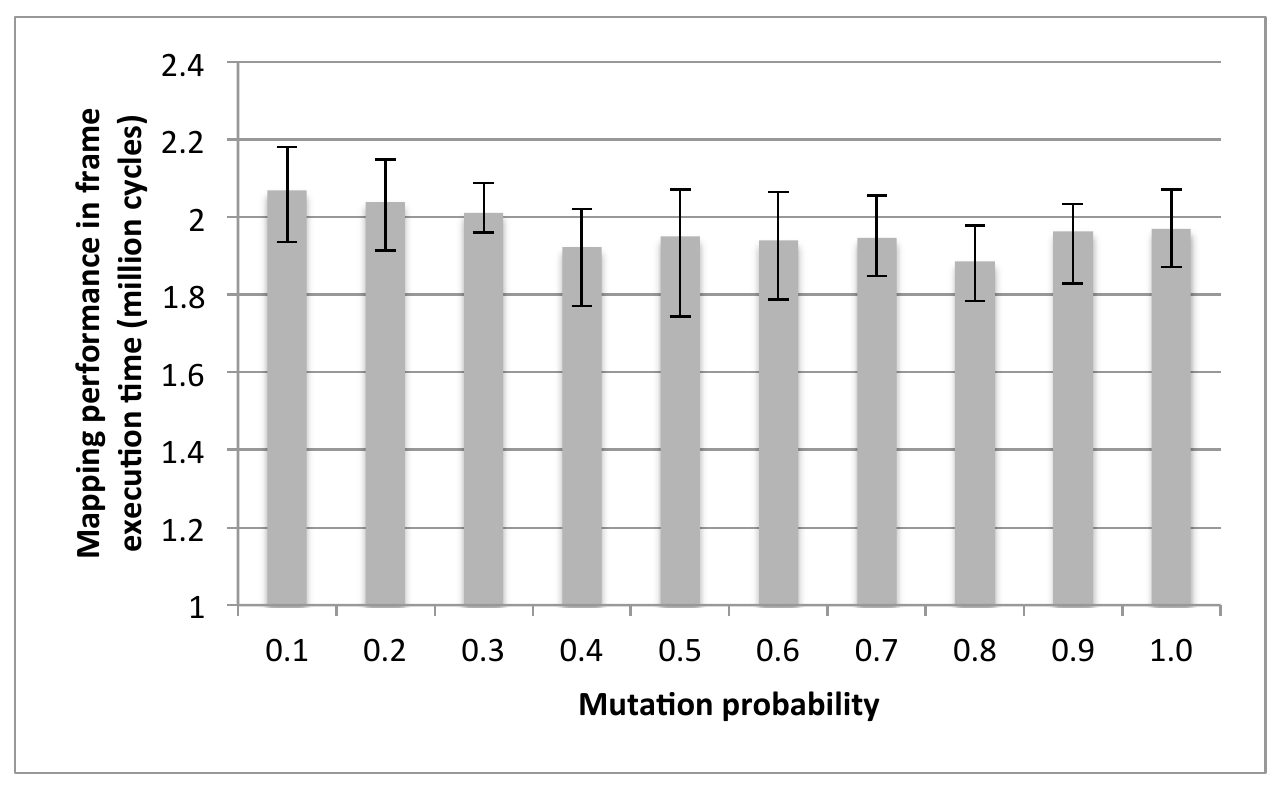}
        }
\subfigure[Algorithm execution time cost (10 executions) of the BEG algorithm with different mutation probabilities]{
            \label{Fig cmutation}
            \includegraphics[width=2.6in]{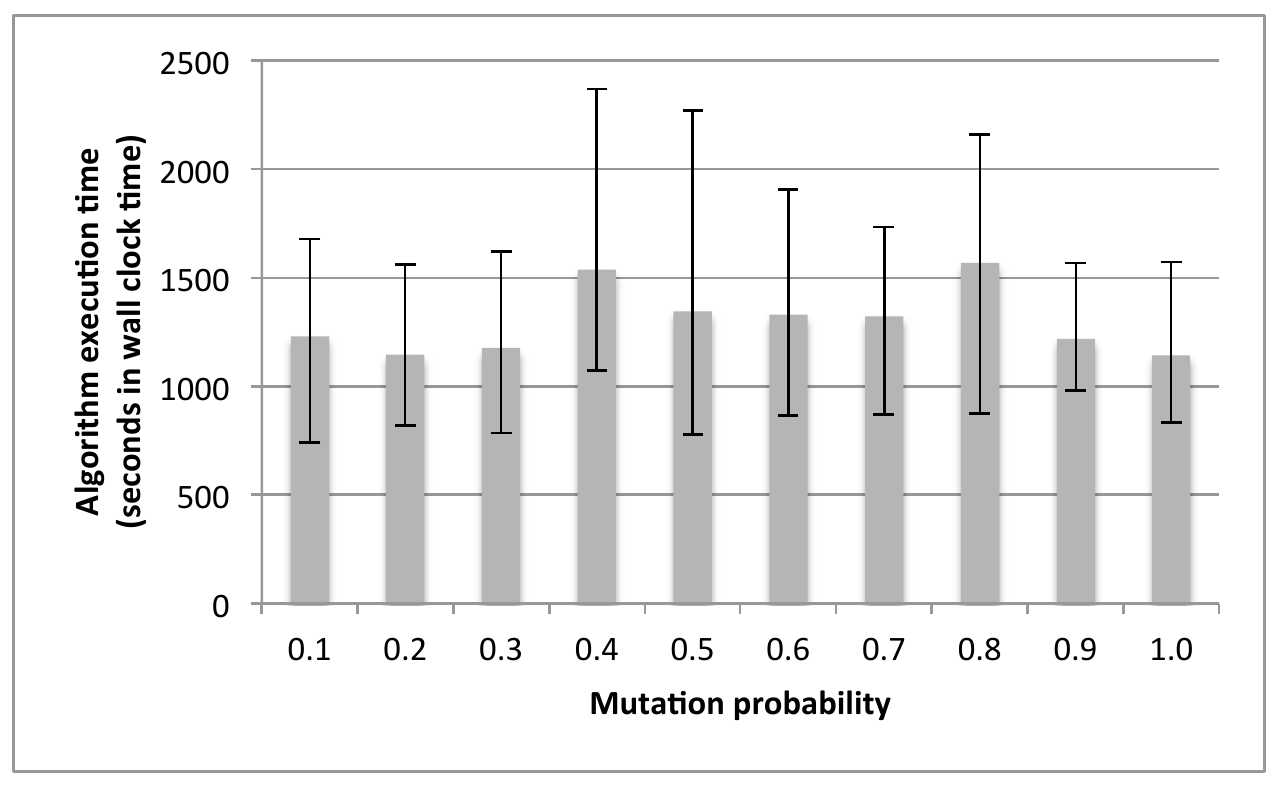}
        }
\caption{Comparison of final mapping performance and algorithm execution time cost of the BEG algorithm with different mutation probabilities.}
\label{figure7}
\end{figure}

To study how the other operators (beside the mutation operator) like the crossover and selection method influence the behaviour of our algorithm, we have applied different operator combinations to BEG. For this experiment (experiment 5), we focus again on the MP3 decoder application, The GA parameters are shown in Table~\ref{table7}. For the crossover operator, the one-point (1p), two-point (2p) and uniform crossover (Uf) are used in this experiment. With regard to the selection approach, the roulette wheel (R), random (Rd) and tournament (T) selection methods have been studied. Figure~\ref{figure6} shows the results of the final mapping quality and algorithm execution time cost for the different operator combinations. In this figure, the x-axis contains the operator combinations. For example, 1pR represents one-point crossover and roulette wheel selection. In Figure~\ref{Fig poperator}, each bar represents the average mapping performance of the final mapping over 10 executions. From Figure~\ref{figure6}, we can see that the two-point crossover besides the one-point crossover and the random selection besides the roulette wheel selection also work well for our BEG algorithm. However, the uniform crossover and tournament selection show poorer results for our BEG algorithm. For the algorithm execution cost shown in Figure~\ref{Fig coperator}, a trend can be observed that shows that higher quality final mappings generally need more time for our algorithm, irrespective of what operators are used.

In the last experiment (experiment 6), the impact of the mutation probability in our BEG algorithm is investigated. In a general GA, the probability of crossover and mutation should be well-tuned to the problem at hand as it may greatly influence the convergence speed and the exploration time of the algorithm. However, as the crossover operator is not the main focus of this paper, we will not change the crossover probability in this experiment. The mutation probability changes from 0.1 to 1.0. The other parameters for the BEG algorithm can again be found in Table~\ref{table7}. The results are shown in Figure~\ref{figure7} where Figure~\ref{Fig pmutation} shows the final mapping quality derived for a specific mutation probability and Figure~\ref{Fig cmutation} gives the corresponding algorithm execution time cost. In Figure~\ref{Fig pmutation}, we can notice that when the mutation probability is small (below 0.4), the BEG algorithm performs like a general genetic algorithm (EG) with respect to the final mapping quality. However, the algorithm execution cost is less than a general GA (EG takes 3,217 seconds on average) which can be seen in Figure~\ref{Fig cmutation}. On the other hand, if the mutation probability is very high (higher than 0.8), our BEG algorithm also yields less high-quality results even though the algorithm execution cost is reduced (because of a faster convergence) compared with BEG with a lower mutation probability. The reason might be that the algorithm will get stuck in the mapping solution space that only contains the mappings with the best makespan and processor workload balance, and the algorithm may not have the chance to explore mappings which are slightly worse in terms of makespan and workload balance. For this particular MP3 task mapping problem, our algorithm can find good results  when the mutation probability is between 0.4 and 0.8. Notice that, even though the mutation probability of our BEG algorithm is fixed in this paper, some adaptive probability adjusting strategies~\cite{Thierens02,Serpell10} can be applied to further optimize our algorithm.

\section{Related Research}
In recent years, much research has been performed in the area of task mapping for embedded systems. ~\cite{Singh13} gives a nice survey of the existing mapping methodologies.

In the context of static mapping performance optimization, some classic algorithms such as Simulated Annealing (SA)~\cite{Orsila07,Lin05}, Genetic Algorithm (GA)~\cite{Choi12,Alexandrescu11,Page10}, Tabu Search~\cite{Manolache08} and Integer Linear Programming (ILP)~\cite{Javaid09} have been proposed. Among these algorithms, the GA is considered to be a good mapping algorithm because it can obtain a good result in a short time period~\cite{Braun01}. There are different forms of GAs that can be used to obtain a better solution. For instance, ~\cite{Wen11} proposes a heuristic-based hybrid genetic-variable neighborhood search algorithm for guiding the search process and ~\cite{Page10} uses eight heuristics to initialize the GA population for getting better solutions. Alexandrescu et al. ~\cite{Alexandrescu11} propose a GA with a 3-Step Mutation which aims at increasing the solution's convergence rate by using a combination of methods to mutate a chromosome. In contrast to these GAs, our domain-knowledge guided GA is proposed to solve the large scale task mapping problems on the heterogeneous MPSoC systems where the computation and communication cost of tasks and resource contention in the system are carefully considered in the evolution process. 

In our approach, a simulator is used to evaluate the fitness of each chromosome which greatly increases the total evolution time. Consequently, design space pruning techniques need to be considered in our work. In this field, the most recently related work is from~\cite{Thompson13} where the system-level design space is implicitly pruned by exploiting domain knowledge in their GA-based DSE. In contrast to our work, however, the work of \cite{Thompson13} only deals with homogeneous systems and enriched their GA  with a "mapping distance"  based crossover operator. Some other approaches, for example~\cite{Ascia07,Mariani10,Palermo09}, perform design space pruning via meta-model assisted optimization, which combines simple and approximate models with more expensive simulation techniques. Another class of design space pruning is based on hierarchical DSE (e.g.,~\cite{Erbas06,Kim06,Zai10}). In these approaches, DSE is first performed using analytical or symbolic models to quickly find the interesting parts in the design space, after which simulation-based DSE is performed to more accurately search for the optimal design points.

\section{Conclusion}
The large scale task mapping problem is hard to solve especially when the communication between tasks also needs to be considered. Even though genetic algorithms have a proven track record in solving such problems, these algorithms still need to be carefully designed in order to obtain high-quality solutions in an acceptable time. In this paper, we have proposed a bias-elitism genetic (BEG) algorithm where the mutation operator has been optimized for our task mapping problem. More specifically, we have added domain-specific heuristics as well as a Minimum Completion Time heuristic to the mutation operator. In addition, the selection method in our genetic algorithm has also been tailored for the purpose of finding a good mapping in a short time period. In various experiments, different state-of-the-art algorithms have been compared to our BEG algorithm.  These experiment results clearly confirm the effectiveness of our algorithm.




%
\bibliographystyle{IEEEtran}
\bibliography{ga_dse} 

\begin{thebibliography}{10}

\bibitem{Alexandrescu11}
A.~Alexandrescu, I.~Agavriloaei, and M.~Craus.
\newblock A genetic algorithm for mapping tasks in heterogeneous computing
  systems.
\newblock In {\em System Theory, Control, and Computing (ICSTCC), 2011 15th
  International Conference on}, pages 1--6, 2011.

\bibitem{Armstrong98}
R.~Armstrong, D.~Hensgen, and T.~Kidd.
\newblock The relative performance of various mapping algorithms is independent
  of sizable variances in run-time predictions.
\newblock In {\em In 7th IEEE Heterogeneous Computing Workshop}, pages 79--87,
  1998.

\bibitem{Ascia07}
G.~Ascia, V.~Catania, A.~G.~D. Nuovo, M.~Palesi, and D.~Patti.
\newblock Efficient design space exploration for application specific
  systems-on-a-chip.
\newblock {\em Journal of Systems Architecture}, 53(10):733 -- 750, 2007.

\bibitem{Braun01}
T.~D. Braun, H.~J. Siegel, N.~Beck, L.~L. B\"{o}l\"{o}ni, M.~Maheswaran, A.~I.
  Reuther, J.~P. Robertson, M.~D. Theys, B.~Yao, D.~Hensgen, and R.~F. Freund.
\newblock A comparison of eleven static heuristics for mapping a class of
  independent tasks onto heterogeneous distributed computing systems.
\newblock {\em J. Parallel Distrib. Comput.}, 61(6):810--837, June 2001.

\bibitem{JCII11}
J.~Castrillon, R.~Leupers, and G.~Ascheid.
\newblock Maps: Mapping concurrent dataflow applications to heterogeneous
  mpsocs.
\newblock {\em IEEE Trans.on Industrial Informatics}, PP(99):1, 2011.

\bibitem{Choi12}
J.~Choi, H.~Oh, S.~Kim, and S.~Ha.
\newblock Executing synchronous dataflow graphs on a spm-based multicore
  architecture.
\newblock In {\em Proceedings of the 49th Annual Design Automation Conference},
  DAC '12, pages 664--671, New York, NY, USA, 2012. ACM.

\bibitem{Dey02}
S.~Dey and S.~Majumder.
\newblock Task allocation in heterogeneous computing environment by genetic
  algorithm.
\newblock In {\em Distributed Computing}, volume 2571 of {\em Lecture Notes in
  Computer Science}, pages 348--352. Springer Berlin Heidelberg, 2002.

\bibitem{Erbas06}
C.~Erbas, S.~Cerav-Erbas, and A.~Pimentel.
\newblock Multiobjective optimization and evolutionary algorithms for the
  application mapping problem in multiprocessor system-on-chip design.
\newblock {\em Evolutionary Computation, IEEE Transactions on}, 10(3):358--374,
  2006.

\bibitem{Javaid09}
H.~Javaid and S.~Parameswaran.
\newblock A design flow for application specific heterogeneous pipelined
  multiprocessor systems.
\newblock In {\em Proceedings of the 46th Annual Design Automation Conference},
  DAC '09, pages 250--253, New York, NY, USA, 2009. ACM.

\bibitem{Zai10}
Z.~J. Jia, A.~Pimentel, M.~Thompson, T.~Bautista, and A.~Nunez.
\newblock Nasa: A generic infrastructure for system-level mp-soc design space
  exploration.
\newblock In {\em Embedded Systems for Real-Time Multimedia (ESTIMedia), 2010
  8th IEEE Workshop on}, pages 41--50, 2010.

\bibitem{Kah74}
G.~Kahn.
\newblock The semantics of a simple language for parallel programming.
\newblock In {\em Information processing}, pages 471--475. North Holland,
  Amsterdam, Aug 1974.

\bibitem{Kim06}
J.~Kim and M.~Orshansky.
\newblock Towards formal probabilistic power-performance design space
  exploration.
\newblock In {\em Proceedings of the 16th ACM Great Lakes symposium on VLSI},
  GLSVLSI '06, pages 229--234, New York, NY, USA, 2006. ACM.

\bibitem{KumarISCA04}
R.~Kumar, D.~M. Tullsen, P.~Ranganathan, N.~P. Jouppi, and K.~I. Farkas.
\newblock Single-isa heterogeneous multi-core architectures for multithreaded
  workload performance.
\newblock In {\em Proceedings of the 31st annual international symposium on
  Computer architecture}, ISCA '04, pages 64--75, Washington, DC, USA, 2004.
  IEEE Computer Society.

\bibitem{Lin05}
L.-Y. Lin, C.-Y. Wang, P.-J. Huang, C.-C. Chou, and J.-Y. Jou.
\newblock Communication-driven task binding for multiprocessor with latency
  insensitive network-on-chip.
\newblock In {\em Proceedings of the ASP-DAC 2005}, volume~1, pages 39--44 Vol.
  1, 2005.

\bibitem{Manolache08}
S.~Manolache, P.~Eles, and Z.~Peng.
\newblock Task mapping and priority assignment for soft real-time applications
  under deadline miss ratio constraints.
\newblock {\em ACM Trans. Embed. Comput. Syst.}, 7(2):19:1--19:35, Jan. 2008.

\bibitem{Mariani10}
G.~Mariani, A.~Brankovic, G.~Palermo, J.~Jovic, V.~Zaccaria, and C.~Silvano.
\newblock A correlation-based design space exploration methodology for
  multi-processor systems-on-chip.
\newblock In {\em Proceedings of the 47th Design Automation Conference}, DAC
  '10, pages 120--125, New York, NY, USA, 2010. ACM.

\bibitem{OrsilaThesis}
H.~Orsila.
\newblock {\em Optimizing Algorithms for Task Graph Mapping on Multiprocessor
  System on Chip}.
\newblock PhD thesis, Tampere University of Technology, Finland, 2011.

\bibitem{Orsila07}
H.~Orsila, T.~Kangas, E.~Salminen, T.~D. H\"{a}m\"{a}l\"{a}inen, and
  M.~H\"{a}nnik\"{a}inen.
\newblock Automated memory-aware application distribution for multi-processor
  system-on-chips.
\newblock {\em J. Syst. Archit.}, 53(11):795--815, Nov. 2007.

\bibitem{Page10}
A.~J. Page, T.~M. Keane, and T.~J. Naughton.
\newblock {Multi-heuristic dynamic task allocation using genetic algorithms in
  a heterogeneous distributed system.}
\newblock {\em Journal of parallel and distributed computing}, 70(7):758--766,
  July 2010.

\bibitem{Palermo09}
G.~Palermo, C.~Silvano, and V.~Zaccaria.
\newblock Respir: A response surface-based pareto iterative refinement for
  application-specific design space exploration.
\newblock {\em Computer-Aided Design of Integrated Circuits and Systems, IEEE
  Transactions on}, 28(12):1816--1829, 2009.

\bibitem{PimentelEP06}
A.~D. Pimentel, C.~Erbas, and S.~Polstra.
\newblock A systematic approach to exploring embedded system architectures at
  multiple abstraction levels.
\newblock {\em IEEE Trans. Computers}, 55(2):99--112, 2006.

\bibitem{Serpell10}
M.~Serpell and J.~E. Smith.
\newblock Self-adaptation of mutation operator and probability for permutation
  representations in genetic algorithms.
\newblock {\em Evol. Comput.}, 18(3):491--514, Sept. 2010.

\bibitem{Singh13}
A.~K. Singh, M.~Shafique, A.~Kumar, and J.~Henkel.
\newblock Mapping on multi/many-core systems: survey of current and emerging
  trends.
\newblock In {\em Proceedings of the 50th Annual Design Automation Conference},
  DAC '13, pages 1:1--1:10, New York, NY, USA, 2013. ACM.

\bibitem{Thierens02}
D.~Thierens.
\newblock Adaptive mutation rate control schemes in genetic algorithms.
\newblock In {\em Evolutionary Computation, 2002. CEC '02. Proceedings of the
  2002 Congress on}, volume~1, pages 980--985, 2002.

\bibitem{Thompson13}
M.~Thompson and A.~D. Pimentel.
\newblock Exploiting domain knowledge in system-level mpsoc design space
  exploration.
\newblock {\em J. Syst. Archit.}, 59(7):351--360, Aug. 2013.

\bibitem{Wen11}
Y.~Wen, H.~Xu, and J.~Yang.
\newblock A heuristic-based hybrid genetic-variable neighborhood search
  algorithm for task scheduling in heterogeneous multiprocessor system.
\newblock {\em Information Sciences}, 181(3):567 -- 581, 2011.

\end{thebibliography}

\end{document}